\documentclass[aps,prl,twocolumn,amsmath,amssymb, superscriptaddress]{revtex4-2}
\usepackage{graphicx}
\graphicspath{{Figures/}}
\usepackage{subfigure}
\usepackage{epsfig}
\usepackage{ulem}
\usepackage{dcolumn}
\usepackage{bm}
\usepackage{hyperref}
\hypersetup{colorlinks=true, citecolor=blue, urlcolor=blue, linkcolor=blue}
\renewcommand{\v}[1]{{\boldsymbol{#1}}}

\newcommand{\imth}{\hspace{1pt}\mathrm{i}\hspace{1pt}}
\newcommand{\dif}{\mathrm{d}}

\newcommand{\E}[1]{{\mathrm{e}^{ #1}}}

\newcommand{\p}{\partial}

\newcommand{\ra}{\rightarrow}
\newcommand{\s}{{\sigma}}

\def\beq{\begin{equation}}
\def\eeq{\end{equation}}
\def\bald{\begin{aligned}}
\def\eald{\end{aligned}}
\def\bea{\begin{eqnarray}}
\def\eea{\end{eqnarray}}

\def\inc#1{\left(#1\right)}
\def\Inc#1{\left[#1\right]}

\def\ket#1{\left|#1\right\rangle}
\def\avg#1{\left\langle#1\right\rangle}
\def\braket#1#2{\left\langle #1\right|\left.#2\right\rangle}
\def\expectation#1#2#3{\left\langle #1\left| #2 \right| #3\right\rangle}

\def\Eq#1{Eq.~(\ref{#1})}
\def\Fig#1{Fig.~\ref{#1}}

\begin{document}
\title{Antiferromagnetism induced by electron-phonon-coupling}
\author{Xun Cai}
\affiliation{Institute for Advanced Study, Tsinghua University, Beijing, 100084, China.}
\author{Zi-Xiang Li}
\affiliation{Department of Physics, University of California, Berkeley, CA 94720, USA.}
\affiliation{Materials Sciences Division, Lawrence Berkeley National Laboratory, Berkeley, CA 94720, USA.}
\author{Hong Yao}
\email{yaohong@tsinghua.edu.cn}
\affiliation{Institute for Advanced Study, Tsinghua University, Beijing, 100084, China.}
\affiliation{State Key Laboratory of Low Dimensional Quantum Physics, Tsinghua University, Beijing 100084, China.}

\begin{abstract}
Antiferromagnetism (AF) such as Neel ordering is often closely related to Coulomb interactions such as Hubbard repulsion in two-dimensional (2D) systems.
Whether Neel AF ordering in 2D can be dominantly induced by electron-phonon couplings (EPC) has not been completely understood.
Here, by employing numerically-exact sign-problem-free quantum Monte Carlo (QMC) simulations, we show that optical Su-Schrieffer-Heeger (SSH) phonons with frequency $\omega$ and EPC constant $\lambda$ can induce AF ordering for a wide range of phonon frequency $\omega>\omega_c$.
For $\omega<\omega_c$, a valence-bond-solid (VBS) order appears and there is a direct quantum phase transition between VBS and AF phases at $\omega_c$.
The phonon mechanism of the AF ordering is related to the fact that SSH phonons directly couple to electron hopping whose second-order process can induce an effective AF spin exchange.
Our results shall shed new lights to understanding AF ordering in correlated quantum materials.
\end{abstract}
\date{\today}

\maketitle
{\bf Introduction:} Electron-phonon coupling (EPC) exists ubiquitously in quantum materials. Moreover, it plays a crucial role in driving various exotic quantum phenomena, including
charge-density wave (CDW) order \cite{Peierls1955book,Grunner1988RMP}, Su-Schrieffer-Heeger topological state \cite{originalSSH, SSH-review}, and, most notably, BCS superconductivity (SC) \cite{BCSoriginal,Schrieffer1964book}.
Since EPC normally induces an effective attraction between electrons, it has been well understood theoretically that EPC induces charge-density-wave, bond-density-wave, or conventional superconductivity in quantum systems, which was further illustrated in recent works \cite{Kivelson2018PRB-MEbreakdown, Kivelson2019PRB-HolsteinPseudogap,  ZXLi2019PRB-HolsteinFrustration, Scalettar2018PRL, Scalettar2019PRL-holstein, Hohenadler2019PRL-holstein,  Scalettar2019EPC, CubicHolsteinLangevinQMC2020PRB, Scalettar2020arx-FlatBandHolstein, Louie2019EPC,Xiang2020SigmaEPC}.
Nonetheless, the role of EPC in driving antiferromagnetism (AF) and unconventional SC in strongly correlated systems has been under debate since it is widely believed that repulsive Coulomb interactions between electrons are essential in developing AF and unconventional SC (including high-temperature SC) \cite{Anderson1987RVB,Kivelson2003RMP, Anderson2004-vanillaRVB, Wen2006RMP,Scalapino2012RMP,DHLee2013AFM}.

In the past many years, increasing experimental and theoretical studies suggest that EPC can be an essential ingredient in understanding high-temperature SC, including cuprates \cite{Lanzara2001Cuprate,ZXShen2002PhilosophicalReview, ZXShen2004PRL, Nagaosa2004PRL, ZXShen2005PRL-phonon-cuprate, Nagaosa2005ARPESreview, ZXShen2005PRL, Cooper2005PRL, Davis2006Nature-phonon-no-effect, DevereauxCuprate2010PRB, ZXShen2017Science, ZXShenARPES,ZaanenReviewCuprate,Zhang2016Phonon,Xue2016cuprate, Kivelson2019ThermalHall,PhysRevLett.105.257001,PhysRevB.98.035102} and iron-based superconductors \cite{Wang2012CPL,ZXShen2014Nature-replicabandFeSe, ZXLi2016FeSe, Johnston2016PRB,DLFeng2017FeSe, JDGuo2017FeSe,Millis2017FeSe,Moses2018SciAdv-FeSe, ZXLi2019PRB-smallmomentum, RuiPeng2020SciAdv,Hoffman2017FeSeReview,DHLee2018Review}, and in driving exotic orders such as pair-density-wave \cite{ZYHan2020PRL},
raising renewed interests in studying the role of EPC in correlated quantum systems.
Since unconventional SC is often closely related to antiferromagnetism  \cite{Anderson1987RVB,Kivelson2003RMP, Anderson2004-vanillaRVB, Wen2006RMP,Scalapino2012RMP,DHLee2013AFM}, it is natural to ask whether EPC can play any essential role in driving AF ordering. 
Although various previous works have studied the competition between AF order dominantly induced by the Hubbard interactions and other types of orders such as CDW induced by EPC \cite{Assaad1996PRL-dwave, Scalapino1997PRB, Sawatzky2004EPC, Millis2007EPC, Johnston2013PRB, Imada2017EPC, Serella2019arXiv, Campbell2003PRB, Costa2020arx, Hohenadler2019PRB, Kivelson2020nematic, DHLee2007PRB-fRG-dwave, Hague2007Bipolaron, Wang2015Holstein, Wang2020EPC, ShiweiZhang2020arXiv}, whether AF ordering can be induced dominantly by EPC remains elusive.

\begin{figure}[t]
    \includegraphics[width=0.8\linewidth]{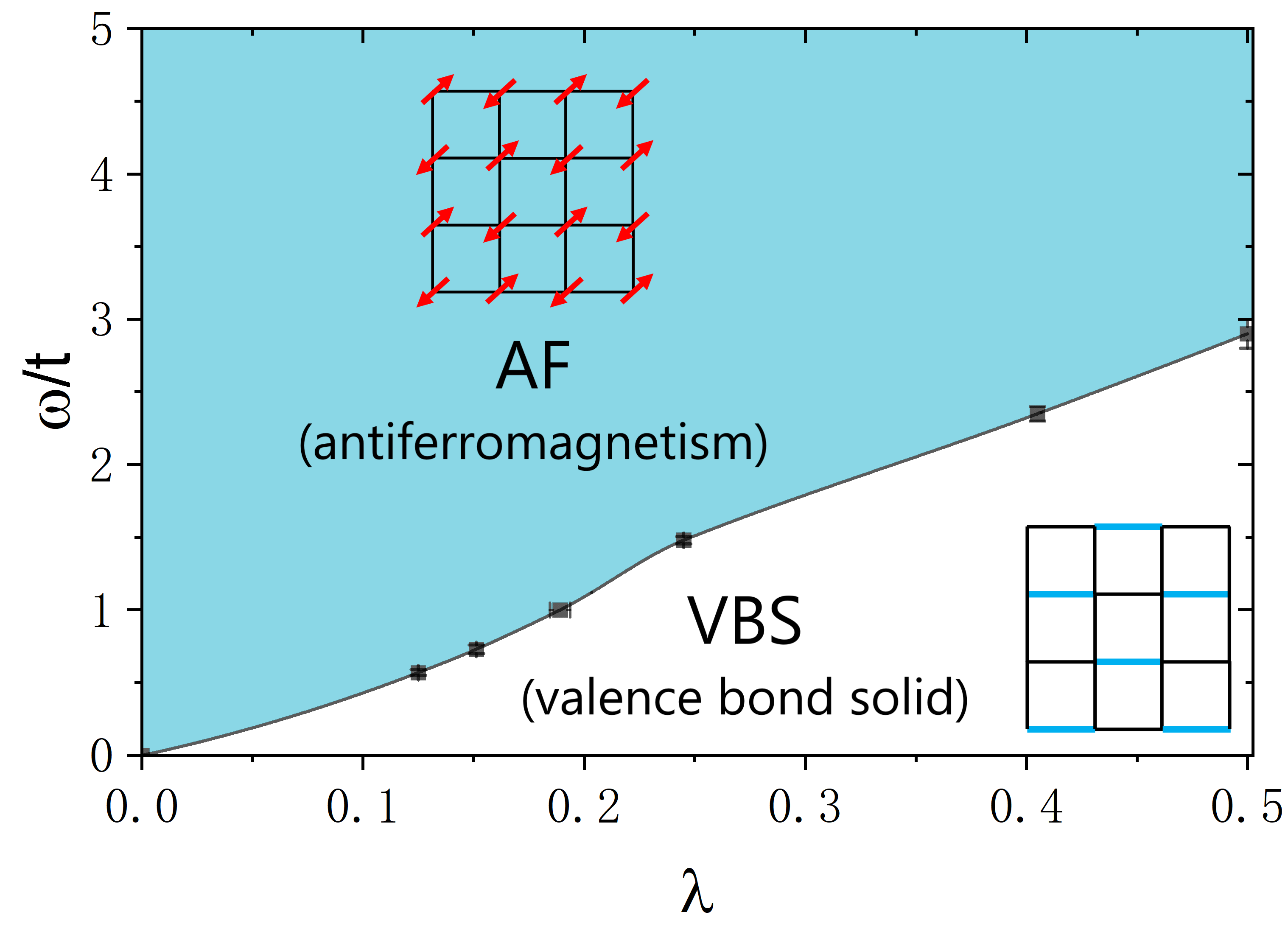}
    \caption{The quantum phase diagram of the square lattice optical SSH model at half filling as a function of dimensionless electron-phonon-coupling (EPC) constant $\lambda$ and phonon frequency $\omega$. The insets depict AF and VBS orders. The results are obtained by large-scale sign-problem-free zero-temperature QMC simulations.}
    \label{FigPhaseDiagram}
\end{figure}

In this Letter, we fill in the gap by convincingly showing that an AF insulator can be dominantly induced by phonons.
Specifically, we systematically study the square lattice Su-Schrieffer-Heeger (SSH) optical phonon model at half filling by performing large-scale quantum Monte Carlo (QMC) simulations \cite{BSS1981PRD,Assaadnote}. The simulations of the model can be rendered sign-problem-free so that we can access large system sizes to reach reliable results \cite{ZXLi2015PRB,ZXLi2016PRL,TXiang2016PRL,Berg2012Science, CJWu2005PRB, Troyer2005PRL,LWang2015PRL} (for a recent review of sign-free QMC, see Ref. \cite{ZXLiQMCreview}). Although acoustic and optical SSH phonon models have been studied by various theoretical and numerical approaches \cite{Lieb1995EPC,Hamano2000JPSJ,Assaad2018SSH,Johnston2020SSH,Scalettar2021PRL, Fradkin1983SSH,Sandvik2003SSH,Bourbonnais2007PRB,Stamp2010SSH, Assaad2012SSH,Sous2018SSH,Hohenadler2015SSH,Hohenadler2020SSH,Weber2021SSH,Sous2020SSH}, it hasn't been shown that AF ordering can be dominantly triggered by SSH phonons.
By performing the first state-of-the-art zero-temperature QMC simulation on the 2D SSH model of optical phonons with frequency $\omega$, we are able to obtain its reliable ground-state phase diagram as a function of $\omega$ and EPC constant $\lambda$, revealing that the AF ordering emerges in a large portion of the phase diagram, as shown in \Fig{FigPhaseDiagram}. 
To the best of our knowledge, it is the first time that an AF insulator is shown, in an numerically-exact way, to be dominantly driven by EPC rather than by Coulomb repulsions between electrons. 
We would like to emphasize that the phonon mechanism of AF ordering is intimately related to the fact that SSH phonons couple to electron hopping whose second-order process can induce an effective AF spin exchange and drive an AF ordering, as we shall explain below.

{\bf Model:} We consider the optical SSH model on the square lattice with the following Hamiltonian
\bea\label{EqOriginalModel}
&&H=-t\sum_{\avg{ij}}(c^\dagger_{i\s}c_{j\s}+h.c.) +\sum_{\avg{ij}}\frac{\hat{P}^2_{ij}}{2M}+\frac{K}{2}\hat{X}^2_{ij} \nonumber\\
&&~~~~~+g\sum_{\avg{ij}}\hat{X}_{ij}(c^\dagger_{i\s}c_{j\s}+h.c.),
\eea
where $\avg{ij}$ refers to the bond between nearest-neighbor (NN) sites $i$ and $j$, $c^\dagger_{i\s}$ creates an electron on site $i$ with spin polarization $\s=\uparrow$/$\downarrow$, $\hat{X}_{ij}$ and $\hat P_{ij}$ are the displacement and momentum operators of the optical SSH phonons on the NN bond $\avg{ij}$. The chemical potential $\mu$ is implicit in the Hamiltonian and hereafter we shall focus on the case of half-filling by setting $\mu=0$. Here $t$ is the electron hopping amplitude and SSH phonon frequency is $\omega=\sqrt{K/M}$. The displacement field of SSH phonons is linearly coupled to electron's NN hopping rather than to electron density. The strength of EPC can be characterized by the dimensionless EPC constant $\lambda\equiv\frac{g^2/K}{W}$, where $W=8t$ is the characteristic band width of the square lattice. Hereafter, we set $t=1$ as energy unit and set $K=1$ by appropriately redefining $\hat{X}_{ij}$.

It is worth noting that the optical SSH phonon model at half filling described by \Eq{EqOriginalModel} respects the SO(3)$\otimes$SO(3)$\otimes$$\mathbb{Z}_2$$\otimes$$\mathbb{Z}_2$ symmetry, which is equivalent to O(4) symmetry \cite{CNYang-SCZhang}. Here the first SO(3) refers to spin rotational symmetry, the second SO(3) pseudospin symmetry \cite{SCZhang1990PRL}, the first $\mathbb{Z}_2$ the usual particle-hole symmetry for both spin-up and spin-down electrons ($c_{i\s}\to (-1)^i c^\dag_{i\s}$), and the second $\mathbb{Z}_2$ the particle-hole symmetry for spin-down electrons $(c_{i\downarrow}\to (-1)^i c^\dag_{i\downarrow})$. The pseudospin rotation can transform the CDW order $\frac{1}{N}\sum_i(-1)^i\langle c^\dag_{i\s}c_{i\s}\rangle$ to the SC order $\frac{1}{N}\sum_i \langle c^\dag_{i\uparrow}c^\dag_{i\downarrow}\rangle$ \cite{SCZhang1990PRL}, where $N=L\!\times\! L$ is the system size. The spin-down particle-hole symmetry can transform the usual AF ordering into pseudospin-AF ordering (pseudospin-AF referring to CDW/SC) \cite{FradkinFieldTheory} so that AF and pseudospin-AF order parameters are degenerate.
The Hubbard interaction $H_{U}=U\sum_{i}(n_{i\uparrow}-\frac{1}{2}) (n_{i\downarrow}-\frac{1}{2})$, which breaks the second $\mathbb{Z}_2$ symmetry explicitly, can lift the degeneracy between the AF and pseudospin-AF ordering; AF ordering is favored over pseudospin-AF ordering by any finite (even infinitesimal) Hubbard repulsion $U>0$. Hereafter, we implicitly assume that a weak Hubbard repulsion $U$ is present to break the degeneracy between AF and pseudospin-AF at half-filling. The model $\tilde H=H+H_U$ is dubbed as the Su-Schrieffer-Heeger-Hubbard model.

\begin{figure}[t]
\subfigure{\label{figJAFMratio}\includegraphics[width= 0.48\linewidth]{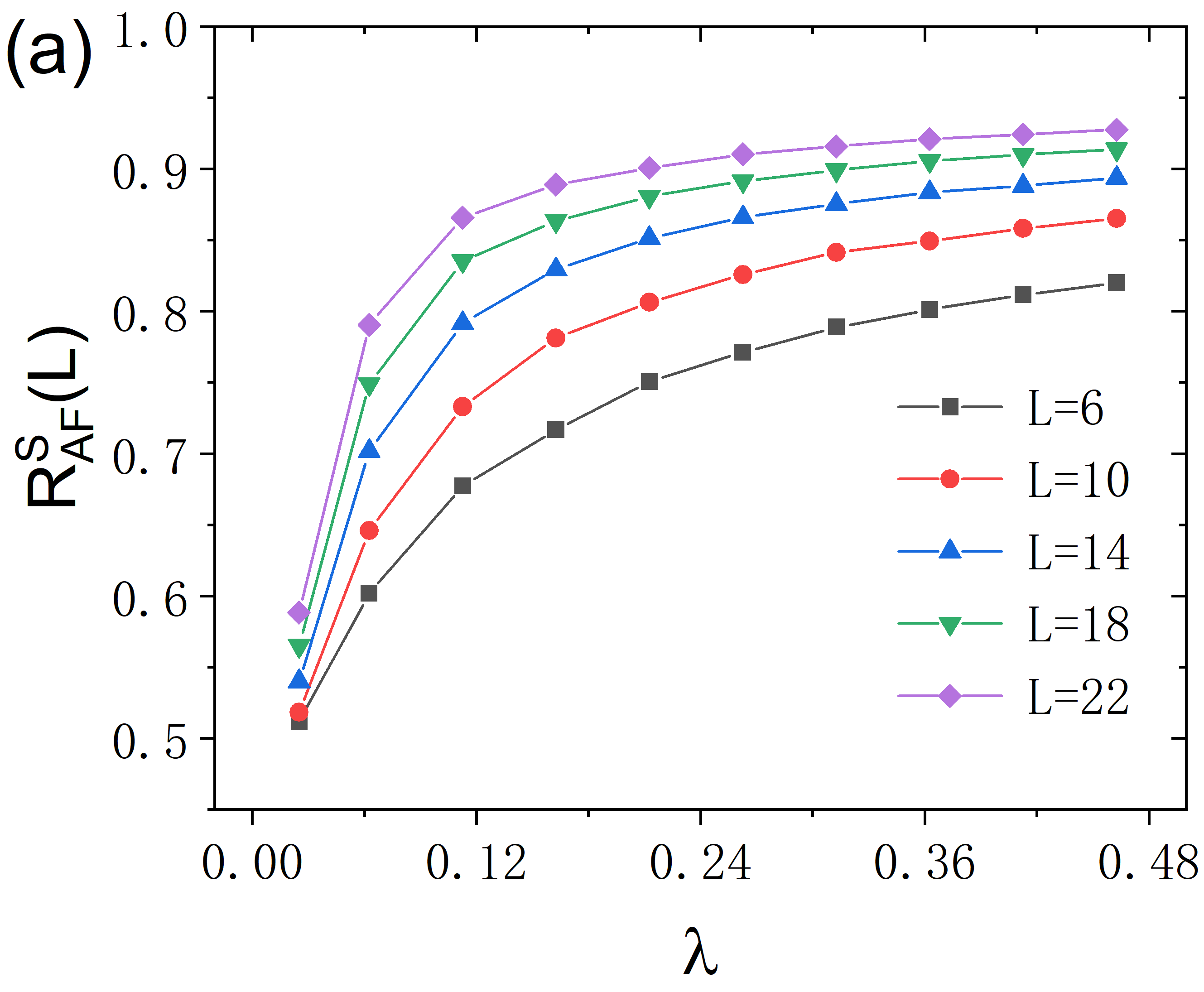}}~~~~	
\subfigure{\label{figJAFMorder}\includegraphics[width= 0.48\linewidth]{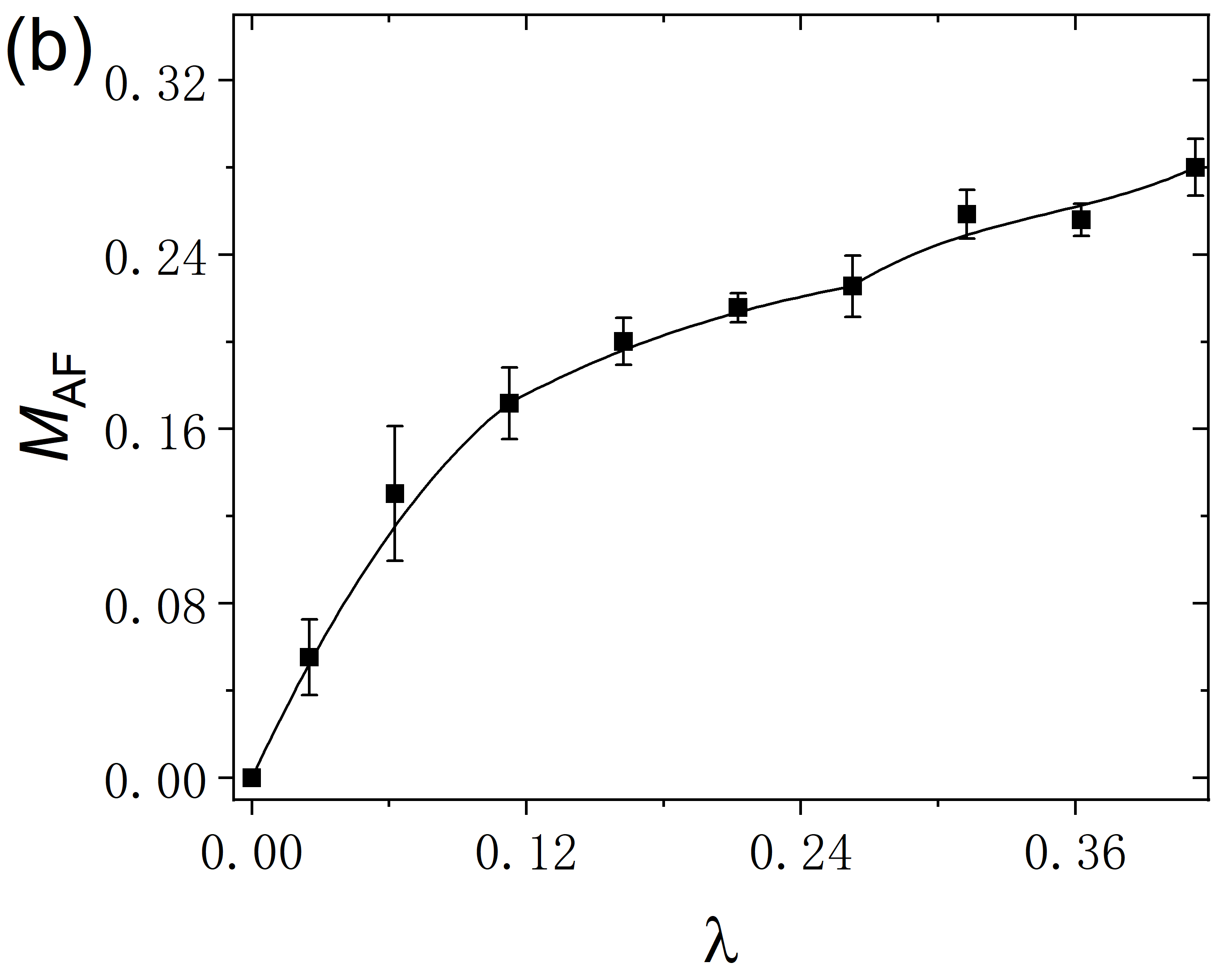}}
\caption{The QMC results of AF correlations in the anti-adiabatic limit ($\omega=\infty$). (a) The AF correlation ratio $R^S_\text{AF}$ as a function of dimensionless EPC constant $\lambda$ for different $L$.
(b) The extrapolated AF order parameter $M_\text{AF}=|\langle\v{S}_i\rangle|$ to the thermodynamic limit ($L\to \infty$) as a function of $\lambda$. In the anti-adiabatic limit, AF ordering occurs for any $\lambda>0$.}
\label{FigJresult}
\end{figure}

The optical SSH model in \Eq{EqOriginalModel} is sign-problem-free so that we can perform large-scale projector QMC simulations to investigate its ground-state phase diagram by varying phonon frequency $\omega$ and EPC constant $\lambda$.
The projector QMC is numerically-exact and is able to study the zero-temperature properties directly. Details of the projector QMC method can be found in the Supplemental Material (SM).
We emphasize that the simulations here are free from the notorious sign problem \cite{ZXLiQMCreview} so that we can study large system size.
To investigate various possible symmetry-breaking orders, we compute the structure factor $S(\v{q},L)\!=\!\frac{1}{N^2}\sum_{i,j}\text{e}^{\text{i}\v{q}\cdot (\v{r_i}-\v{r_j})}\langle \hat O_i\hat O_j\rangle$ of the corresponding order $O$ and evaluate the RG-invariant ratio of the structure factor, namely the correlation ratio, $R^S(L)\!=\!1-\frac{S(\v{Q}+\v{\delta q},L)}{S(\v{Q},L)}$, where $\v{Q}$ refers to the ordering momentum and $\v{\delta q}\!=\!(\frac{2\pi}{L},\frac{2\pi}{L})$ is a minimal momentum shift from $\v{Q}$.
For both Neel AF and staggered VBS ordering, $\v{Q}\!=\!(\pi,\pi)$.
In the thermodynamic limit ($L\!\to\!\infty$), an ordered phase is recognized by $R^S\ra 1$ while a disordered phase features $R^S\ra 0$.
For AF ordering, we further compute the susceptibility ratio $R^\chi(L)\!=\!1-\frac{\chi\inc{\v{Q}+\v{\delta q},L}}{\chi\inc{\v{Q},L}}$, where $\chi(\v{q})$ represents magnetic susceptibility at momentum $\v{q}$, as it has smaller finite-size corrections than the correlation ratio \cite{AssaadPRB91_165108} (see the SM for technical details of evaluating susceptibilities in QMC).

{\bf Results in adiabatic and anti-adiabatic limit:} Integrating out phonons with finite frequency
yields an retarded interactions between electrons.
The retardation effect of EPC plays a central role in driving various novel physics, including SC.
The retardation is usually characterized by the ratio between the phonon frequency $\omega$ and the Fermi energy or band width $W$.
Before performing systematic QMC simulations on the SSH model at a generic finite phonon frequency, we first study the ground-state properties at the adiabatic limit ($\omega=0$) and anti-adiabatic limit ($\omega=\infty$), respectively.

\begin{figure}[t]
\subfigure{\label{figVBSratiog14}\includegraphics[width= 0.49\linewidth]{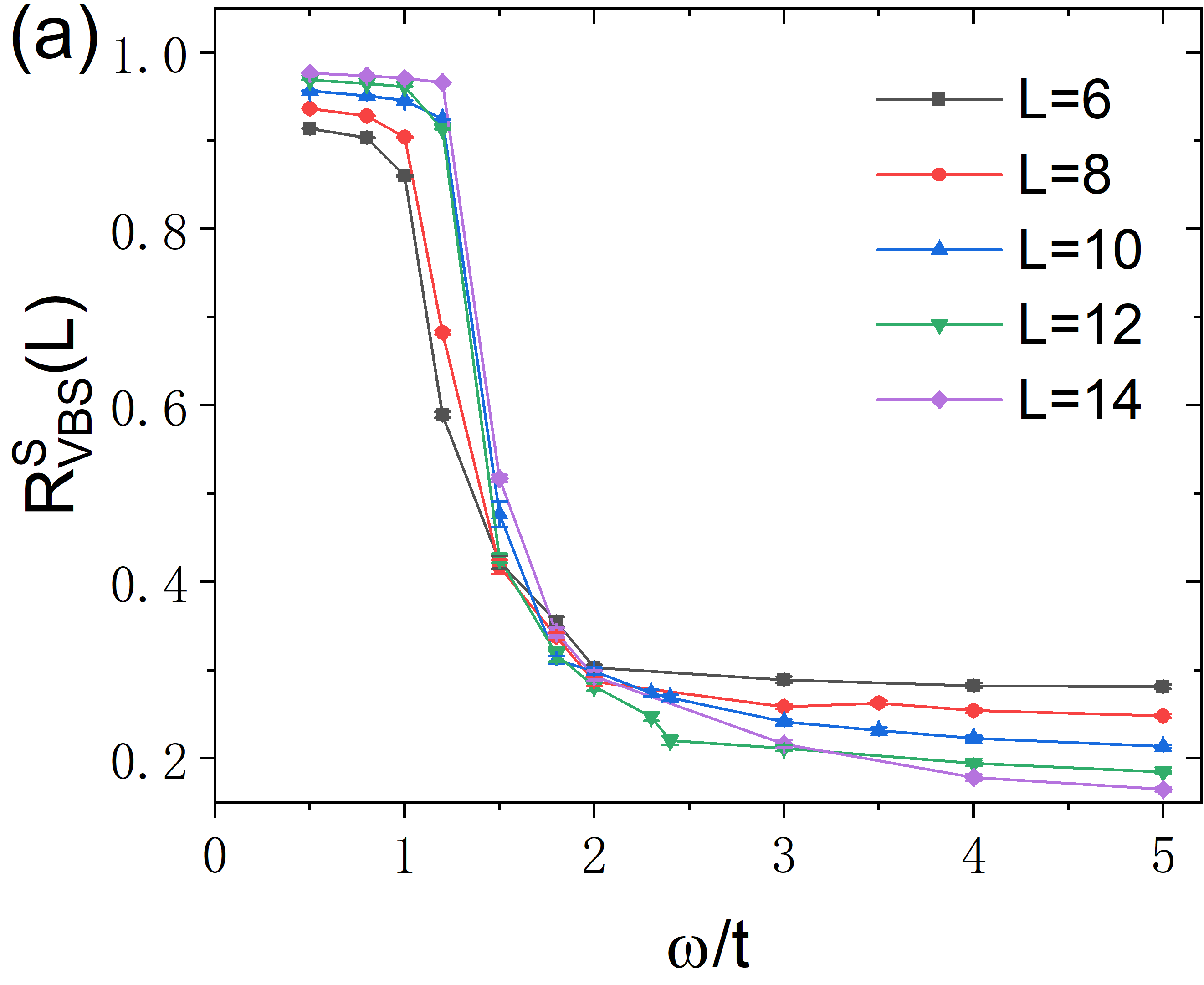}}~~~~
\subfigure{\label{figBondCorrOmega}\includegraphics[width= 0.49\linewidth]{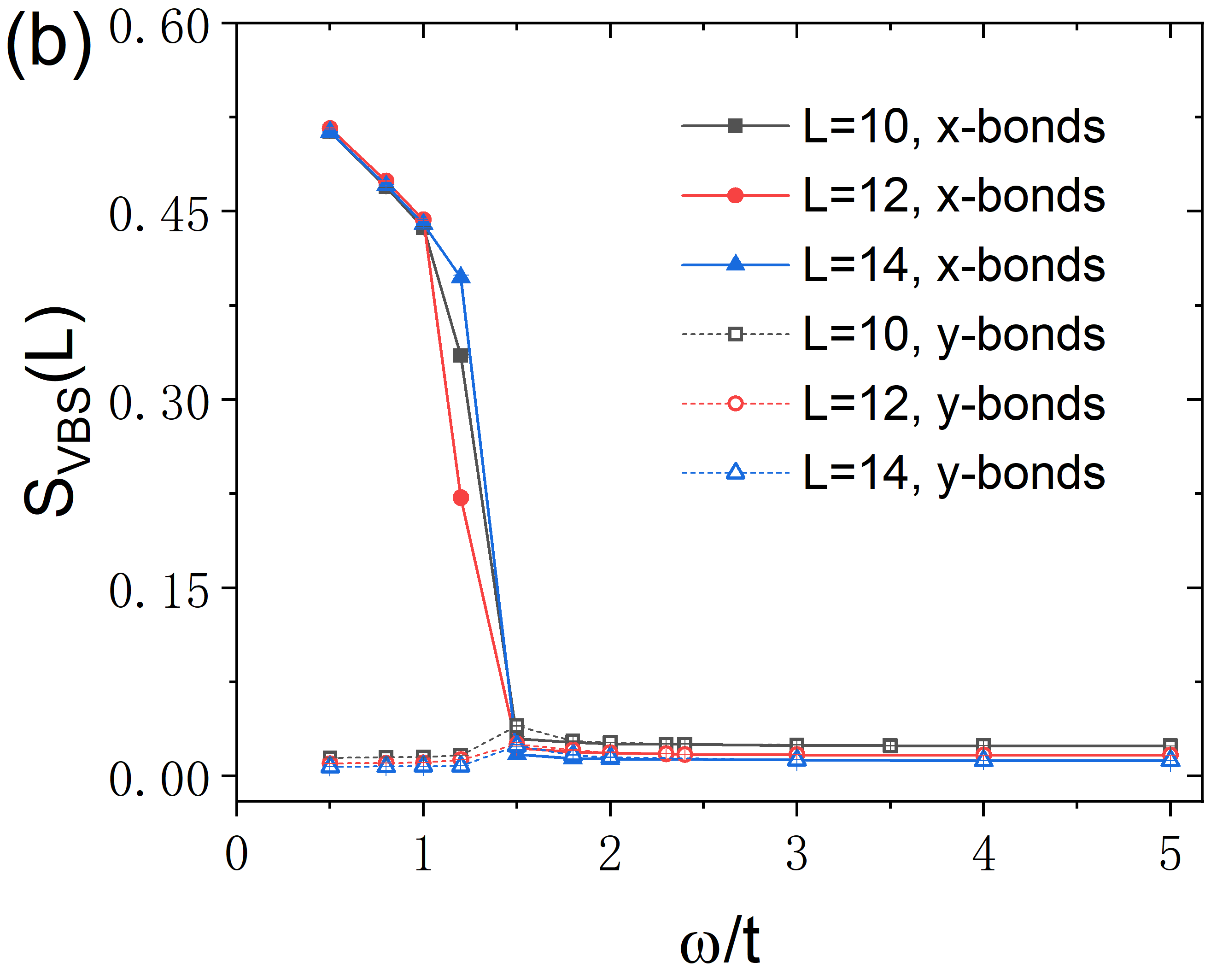}}
\caption{The QMC results of VBS correlations as a function of $\omega$ for $\lambda\approx 0.25$ ($g=1.4$). (a) The crossing of VBS correlation ratio $R_{\mathrm{VBS}}^S$ for different $L$ implies that the VBS transition occurs at $\omega_c\approx 1.5$. (b) The VBS transition at $\omega_c\approx 1.5$ is also observed from the deviation between the structure factor of $x$-bond correlations and $y$-bond correlations.}
\label{FigRatiog14}
\end{figure}

In the adiabatic limit ($\omega=0$), the phonon is static at zero temperature and the exact solution can be obtained by treating the phonon displacement configuration $X_{ij}$ as variational parameters. As electron's bare Fermi surface features a perfect nesting vector $\v{Q}=\inc{\pi,\pi}$, it is natural to expect that the Fermi surface is unstable towards staggered VBS ordering for any finite EPC constant $\lambda$. Indeed, our calculations show that the expectation value of electron hopping on NN bonds alternates in staggered pattern (see the inset of \Fig{FigPhaseDiagram}), which breaks the lattice translational symmetry as well as $\mathbb{C}_4$ rotational symmetry (see the SM for details of the calculations).

In the anti-adiabatic (AA) limit ($\omega=\infty$), the effective electronic interaction mediated by phonons becomes instantaneous, which is proportional to the square of hopping on NN bonds. Consequently, in the AA limit, the original optical SSH model can be reduced to the following effective Hamiltonian
\bea\label{EqHopSquare}
H_\text{AA}=-t\sum_{\avg{ij}}(c^\dagger_{i\s}c_{j\s}\!+\!h.c.) +J\sum_{\avg{ij}}(\v{S}_i\cdot\v{S}_j+ \v{\tilde S}_i\cdot \v{\tilde S}_j),~~~~
\eea
where $J=g^2/K$ is the strength of instantaneous interactions mediated by optical phonons in the AA limit, $\v{S}_i$ and $\v{\tilde S}_i$ are spin and pseudospin operators on site $i$, respectively. Specifically, $\v{S}_i=\frac{1}{2}c^\dag_i \v{\sigma} c_i$ and $\v{\tilde S}_i=\frac{1}{2}\tilde c^\dag_i\v{\sigma}\tilde c_i$, where $c^\dag_i=(c^\dag_{i\uparrow},c^\dag_{i\downarrow})$, $\tilde c^\dag_i=(c^\dag_{i\uparrow},(-1)^i c_{i\downarrow})$, and $\v{\sigma}$ represents the vector of Pauli matrices.
The phonon-mediated interactions include antiferromagnetic spin-exchange interaction, repulsive density-density interaction, and pair hopping terms.
It is worth to emphasize that antiferromagnetic ($J>0$) spin exchange interactions are generated by EPC of SSH phonons, yielding the possibility of AF ordering at half-filling.
By performing QMC simulations on $H_\text{AA}$, we obtained the results of AF correlation ratio and AF order parameter as a function of EPC constant $\lambda=J/W$, as shown in \Fig{FigJresult}.
The AF correlation ratio $R^S_\mathrm{AF}$ monotonically increases with the size $L$ for all studied $\lambda$, as shown in \Fig{figJAFMratio}, indicating that AF ordering occurs for all $\lambda>0$. Furthermore, we obtained the AF order parameter by finite-size scaling to the thermodynamical limit, as shown in \Fig{figJAFMorder}, which reveals that AF order induced by SSH phonons increases with $\lambda$.
Consequently, we conclude that the ground-state of the optical SSH model in the AA limit possesses AF long-range order for any $\lambda>0$. Moreover, it is an AF insulator as its Fermi surface is fully gapped by AF order.

{\bf Antiferromagnetism at finite frequency:} We now study the quantum phase diagram of the SSH model of optical phonons with a generic finite frequency ($0<\omega<\infty$).
Since the ground-state is AF in the AA limit ($\omega\!=\!\infty$) and VBS in the adiabatic limit ($\omega\!=\! 0$), there must be at least one quantum phase transition (QPT) between VBS and AF phases when $\omega$ is varied from $0$ to $\infty$.
Indeed, for a given $\lambda$, our QMC simulations show that there is a direct QPT between AF and VBS phases by varying $\omega$.
For $\lambda \approx 0.25$ ($g=1.4$), the crossing of VBS correlation ratio of different system sizes $L$ implies that the VBS order persists from $\omega=0$ to a critical frequency $\omega_c\approx 1.5$, as shown in \Fig{figVBSratiog14}.
By evaluating dimer correlations on $x$ or $y$ bonds, as shown in \Fig{figBondCorrOmega}, we further verified that the VBS ordering pattern for $0\!<\!\omega\!<\!\omega_c$ is a staggered VBS breaking the lattice $\mathbb{C}_4$ symmetry, similar to the one observed in the adiabatic limit.

\begin{figure}[t]
\begin{minipage}{0.36\linewidth}
\centering
\subfigure{\label{figAFMSusRatiog14}\includegraphics[width=3.8cm]{ 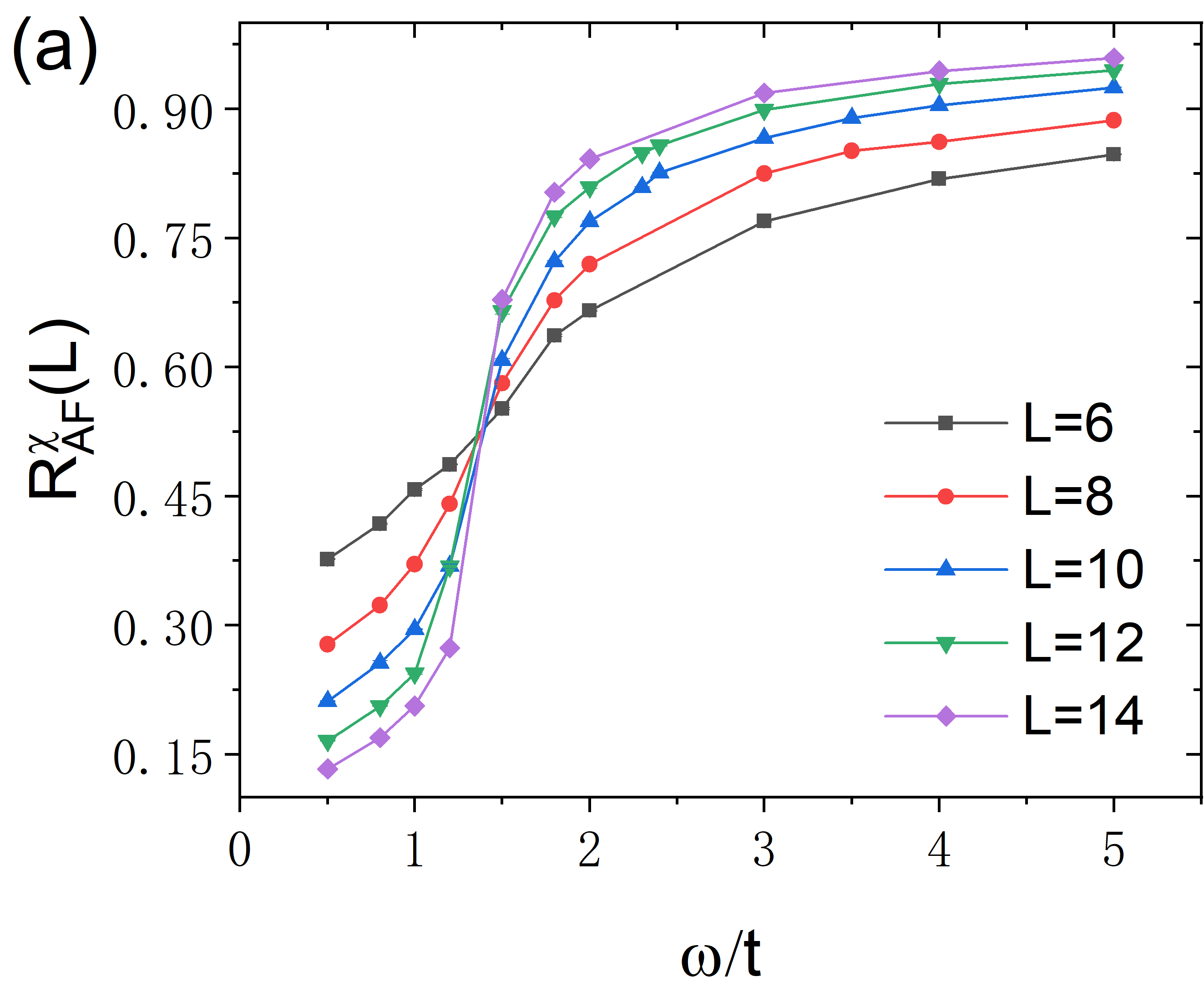}}\\
\subfigure{\label{figAFMSusRatiow10}\includegraphics[width=3.8cm]{ 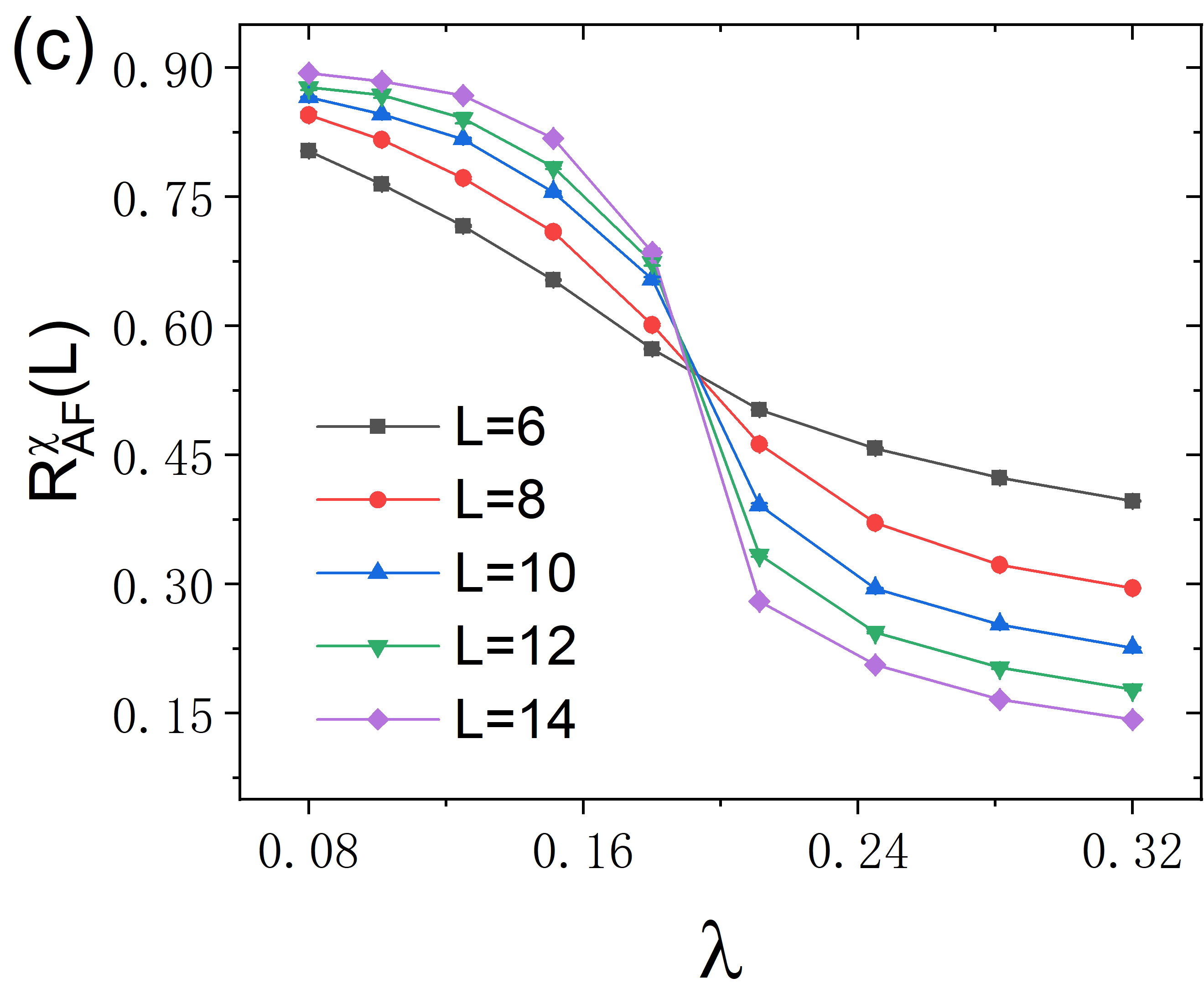}}
\end{minipage}~~~~~~~~
\hspace{0.008 \linewidth}
\begin{minipage}{0.36\linewidth}
\centering
\subfigure{\label{figSpinGapg14}\includegraphics[width=3.8cm]{ 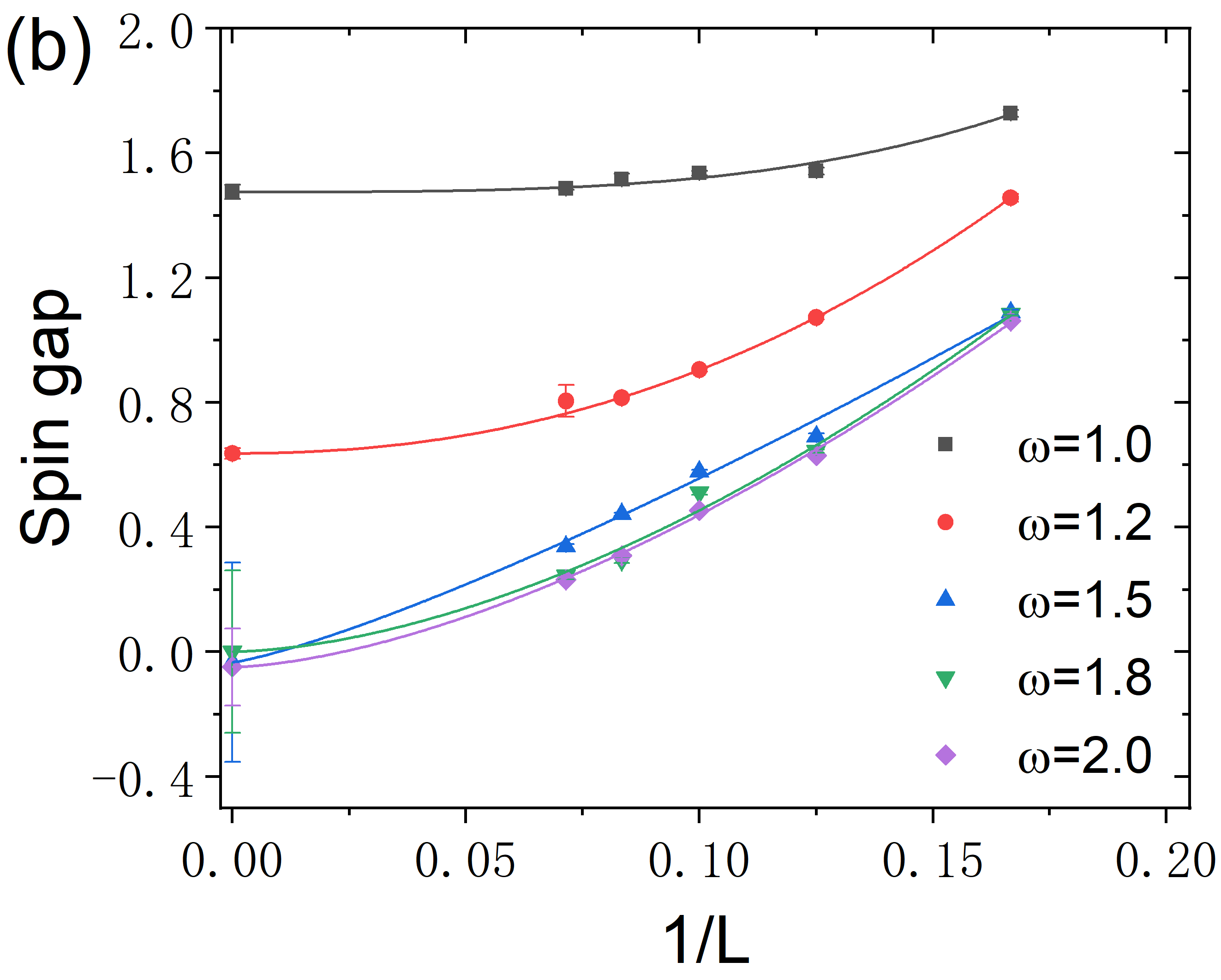}}\\
\subfigure{\label{figAFexchange}\includegraphics[width=4.05cm]{ 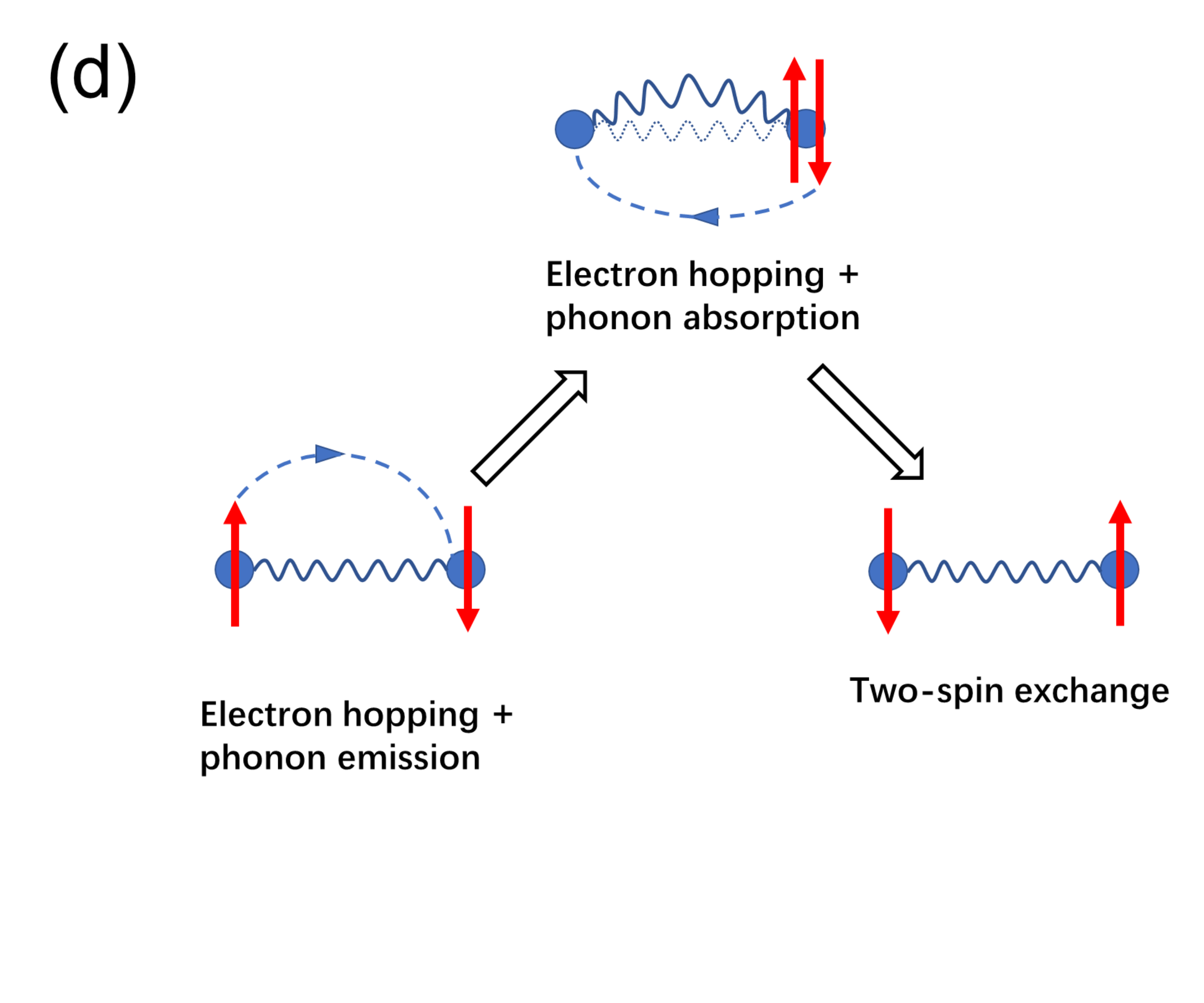}}
\end{minipage}~~~~~~~~
\caption{(a) The QMC results of the AF correlation ratio $R_{\mathrm{AF}}^S$ as the function of $\omega$ for $\lambda\approx 0.25$ ($g=1.4$). The crossing among curves with different $L$ indicates that the AF transition occurs at $\omega=\omega_c\approx 1.5$.
(b) The finite size scaling of the spin gap for $\omega $ near the criticality $\omega_c$.
(c) For fixed $\omega=1.0$, the AF correlation ratio $R_{\mathrm{AF}}^S$ as the function of $\lambda$ and the AF transition occurs at $\lambda_c\approx 0.18$.
(d) The schematic picture of the second-order process of EPC which generates an effective retarded antiferromagnetic spin-exchange interactions.}
\end{figure}

More interestingly, our QMC simulations show that the long-range AF order emerges for $\omega>\omega_c$. Here the critical frequency $\omega_c$ can be accurately extracted from the crossing of AF susceptibility ratio $R_{\mathrm{AFM}}^\chi\inc{L}$ for different $L$. For $\lambda\approx 0.25$ ($g=1.4$), the AF susceptibility ratio $R_{\mathrm{AFM}}^\chi\inc{L}$ displays good crossing near $\omega\approx1.5$, as shown in \Fig{figAFMSusRatiog14}, which implies that AF order develops for $\omega>\omega_c$ with $\omega_c\approx 1.5$. To further verify the AF phase with spontaneous spin-SU(2) rotational symmetry breaking, we compute the spin gap for $\omega$ around  $\omega_c\approx 1.5$, as shown in \Fig{figSpinGapg14}. The spin gap is finite in the VBS regime, but it is extrapolated to zero in the AF regime $\omega>\omega_c$, indicating the emergence of gapless spin-wave excitations as Goldstone modes of spin SU(2) symmetry breaking in the AF phase. Taken together, these results convincingly show that the occurrence of phonon-induced AF long-range order for $\omega>\omega_c$, where $\omega_c$ depends on $\lambda$.

Evidences of AF ordering at $\omega>\omega_c(\lambda)$ are also obtained for various other EPC dimensionless parameters $\lambda$, from weak to strong, as plotted in \Fig{FigPhaseDiagram}. As the critical frequency $\omega_c(\lambda)$ increases monotonically with increasing $\lambda$, for a fixed frequency it is expected that the AF phases should emerge in the regime of $\lambda<\lambda_c$ where $\lambda_c$ is the critical EPC constant. Indeed, for the fixed frequency $\omega=1.0$, AF ordering is observed in the regime of $\lambda<\lambda_c\approx 0.18$ from the crossing of the AF susceptibility ratio for different $L$, as shown in \Fig{figAFMSusRatiow10}. The $(\pi,\pi)$ AF ordering fully gaps out the Fermi surface such that the ground state is an AF insulator for $\lambda<\lambda_c$. As mentioned earlier, the optical SSH model at half-filling respects the $\mathrm{O}(4)$ symmetry, giving rise to the degeneracy between spin AF and pseudospin AF (namely CDW/SC). The degeneracy can be lifted and spin AF is more favored by turning on a weak repulsive Hubbard interaction, as shown in the QMC simulations of models with a weak Hubbard interaction (see the SM for details). 

It is worth to understand heuristically why AF ordering emerges for small $\lambda$. For sufficiently small $\lambda$, one can treat electron-phonon coupling term $g$ as a weak perturbation and the second-order process in $g$ would generate a spin exchange process when the spin polarizations in the NN sites are opposite, as shown in \Fig{figAFexchange}. If the two spins on NN sites are parallel (namely forming a triplet), the exchange process is not allowed. Since this second-order spin-exchange process can gain energy, the spin-exchange interaction is antiferromagnetic. This provides a phonon mechanism to drive AF ordering, which is qualitatively different from the usual AF exchange mechanism of strong Hubbard Coulomb interaction.

Note that AF ordering was not observed in an earlier QMC study of the 2D optical SSH phonon model \cite{Scalettar2021PRL}. There the absence of AF ordering is possibly due to the fact that the QMC simulations were at finite temperature and spin-SU(2) rotational symmetry in 2D cannot be spontaneously broken at any finite temperature.
In contrast, we performed zero-temperature QMC simulations which can directly access properties of the ground state of the 2D phonon model and observe a spontaneous spin-SU(2) symmetry breaking.

We have shown evidences of a direct QPT between the AF and VBS phases.
It is natural to ask if the direct QPT between AF and VBS phases here is first order or continuous. Since AF and VBS phases break totally different symmetries, the QPT between them is putatively first-order in the Landau paradigm although it would be intriguing to explore if a deconfined quantum critical point (DQCP) \cite{Senthil2004Science-DQCP, Senthil2004PRB} occurs in this case.
The phenomena of DQCP have been extensively studied for QPTs between Neel AF and columnar VBS \cite{Sandvik2007DQCP, Kaul2008DQCP, Alet2013DQCP,Shao2016DQCP, Sandvik2013DQCPReview, Wen2016DQCP, Meng2019DQCP,Assaad2019QSH-DQCP, ZXLi2019arXiv-DQCP, Wang2020DQCP, Nahum2020DQCP}. More recently, it has been argued from duality relations that, at such transition point, the SO(5) symmetry might emerge at low energy \cite{Nahum2015DQCP,Nahum2015DQCP,Wang2015Duality, Wang2017DQCP,Metlitski2016Duality, Seiberg2016Duality,Meng2017Duality}.
However, the VBS order in the optical SSH phonon model studied here is the staggered one, for which the VBS $Z_4$ vortex is featureless, namely not carrying a spinon \cite{Levin-Senthil2004, CKXu2011PRB}.
Consequently, a (possibly weak) first-order transition instead of DQCP \cite{Sandvik2010PRB} would be expected for the QPT between AF and staggered VBS phases in the phonon model under study.

{\bf Conclusions and discussions:} We have systematically explored the ground-state phase diagram of the 2D optical SSH model taking account of full quantum phonon dynamics by zero-temperature QMC simulations. Remarkably, from the state-of-the-art numerically-exact simulations, we have shown that the optical SSH phonons can induce a Neel AF order when the phonon frequency is larger than a critical value ($\omega>\omega_c$) or the EPC constant is smaller than a critical value ($\lambda<\lambda_c$). The critical frequency $\omega_c$ can be much smaller than the band width $W$ for weak or moderate EPC constant $\lambda$, which makes the phonon mechanism of AF ordering practically feasible in realistic quantum materials. For instance, for the optical SSH model on the square lattice, we obtained $\omega_c/W \sim 0.1$ when $\lambda\approx 0.15$.

As mentioned above, the role of EPC in understanding the physics of strongly correlated materials, including cuprate and Fe-based high-temperature superconductors, has attracted increasing attentions. We believe that our finding of optical SSH phonon induced AF order may shed new light on understanding the cooperative effects of electronic correlations and EPC on the nature of AF Mott physics. Following this, a natural question to ask is whether such SSH phonon can have crucial effect on unconventional superconductivity arising from doping an AF insulating phase \cite{Anderson1987RVB, Kivelson2003RMP, Anderson2004-vanillaRVB, Wen2006RMP,Scalapino2012RMP,DHLee2013AFM}. In a follow-up work \cite{XunCaiToAppearSoon}, we shall present evidences that quantum SSH optical phonons can substantially enhance the d-wave pairing. We believe that these findings pave an important step to understanding the interplay of EPC and electronic correlations in strongly correlated materials including high-temperature superconductors.

\textit{Acknowledgement}.---We would like to thank Steve Kivelson, Dung-Hai Lee, and Yoni Schattner for helpful discussions. This work is supported in part by the NSFC under Grant No. 11825404 (XC and HY), the MOSTC under Grant Nos. 2016YFA0301001 and 2018YFA0305604 (HY), the CAS Strategic Priority Research Program under Grant No. XDB28000000 (HY), Beijing Municipal Science and Technology Commission under Grant No. Z181100004218001 (HY), and the Gordon and Betty Moore Foundation’s EPiQS under Grant No. GBMF4545 (ZXL).


%

\pagebreak
\widetext
\section{\large Supplemental Material}
  \setcounter{equation}{0}
  \setcounter{figure}{0}
  \setcounter{table}{0}
  \makeatletter
  \renewcommand{\theequation}{S\arabic{equation}}
  \renewcommand{\thefigure}{S\arabic{figure}}
  \renewcommand{\bibnumfmt}[1]{[S#1]}

\subsection{A. The method of projector quantum Monte Carlo}
We employ the method of projector QMC to investigate the ground-state properties of the SSH phonon model at finite frequency described in \Eq{EqOriginalModel} and the effective Hamiltonian described in \Eq{EqHopSquare} in the anti-adiabatic limit. The algorithm is based on the principle that the ground-state expectation value of operator $\hat O$ in the exact ground state $\ket{\psi_G}$ can be computed exactly via projecting a trial wave function $\ket{\psi_T}$ along the imaginary time axis
\beq\label{EqPQMCexpectation}	
\langle\hat{O}\rangle=\frac{\langle\psi_G|\hat{O}|\psi_G\rangle}{\langle\psi_G|\psi_G\rangle} =\lim_{\Theta\ra\infty}\frac{\langle\psi_T|\E{-\Theta H}\hat{O}\E{-\Theta H}|\psi_T\rangle}{\expectation{\psi_T}{\E{-2\Theta H}}{\psi_T}}
\eeq
as long as the trial wave function $\ket{\psi_T}$ has a finite overlap with the true ground state $\ket{\psi_G}$.  
The algorithm is intrinsically unbiased against the choice of $\ket{\psi_T}$ assuming that the trial wave function is not orthogonal to the true ground state, namely $\braket{\psi_T}{\psi_G}\ne 0$, which is generically satisfied for a quantum many-body system with finite size. In this paper, we choose $\ket{\psi_T}$ to be the ground-state wave function of the non-interacting part of the model under consideration. In practical QMC simulations, we set the projection parameter $\Theta$ to a finite but sufficiently large value so that the expectation value $\langle \hat O\rangle$ converges when larger $\Theta$ is considered. We set $\Theta=34/t, 36/t, 38/t,46/t,50/t$ for $L=6,8,10,12,14$ accordingly, each of which has been checked to be large enough for convergence. Similar to finite temperature algorithm, Trotter decomposition is implemented by discretizing $\Theta$ in spacing $\Delta\tau=\Theta/L_\tau$. The Trotter error scales as $\Delta\tau^2$. In this work we choose $\Delta\tau=0.1/t$. The convergence of the discretization has also been checked by comparing results with smaller $\Delta\tau$.

The SSH model in \Eq{EqOriginalModel} becomes quadratic in electron operators for a specific space-time phonon configuration. We compute the electron's Green's function for each phonon configuration, and sample the phonon configuration by Monte Carlo. As for the effective model in the anti-adiabatic limit in \Eq{EqHopSquare}, auxiliary fields are introduced into the Hamiltonian via a SU(2) symmetric Hubbard-Stratonovich decomposition \cite{Assaadnote}. The AF and VBS order parameters for a finite system size $N=L^2$ are given by
\begin{align}
\hat{O}_{\mathrm{AF}}\inc{\v{q}}&=\frac{1}{L^2}\sum_j \E{\imth \v{q}\cdot \v{R}_j} \hat{S}_j^z \label{EqAFMorder}\\
\hat{O}_{\mathrm{VBS}}\inc{\v{q}}&=\frac{1}{L^2}\sum_j \E{\imth \v{q}\cdot \v{R}_j} \inc{\hat{B}_{j,\hat{x}}+\imth \hat{B}_{j,\hat{y}}} \label{EqVBSorder}
\end{align}
where $\hat{B}_{j,\delta}= c^\dagger_{j,\s}c_{j+\delta,\s}+h.c.$ is the kinetic operator on $\delta=\hat{x}, \hat{y}$ bonds. VBS breaks the lattice $\mathbb{Z}_4$ symmetry while AF breaks spin SU(2) rotational symmetry. The structure factors for each order parameter with momentum $\v{Q}$ is defined as $S(\v{Q})=\langle |\hat{O}(\v{Q})|^2\rangle$. For both AF and staggered VBS, the peaked momentum is $\v{Q}=\inc{\pi,\pi}$. The AF susceptibility can be computed as an integration over the imaginary-time interval $\tau_M$
\beq\label{EqSusceptibilityDef}
\chi_{\mathrm{AFM}}\inc{\v{q}}= \int_{\Theta-\tau_M/2}^{\Theta+\tau_M/2} \dif\tau\,\langle\hat{O}_{\mathrm{AFM}} \inc{\v{q},\tau}\hat{O}_{\mathrm{AFM}}\inc{\v{q},0}\rangle.
\eeq
To compute $\chi_{\mathrm{AF}}\inc{\v{q}}$, the imaginary-time interval $\tau_M$ should be large enough such that $\chi$ is converged, while the correlators $\langle \hat{O}\inc{\v{q},\tau}\hat{O}\inc{\v{q},0}\rangle$ should be computed after sufficiently long projection. Thus, in practice we choose $\tau_M\approx\Theta/4$ and integrate over the correlators from $\Theta-\tau_M/2$ to $\Theta+\tau_M/2$.

\subsection{B. Notes on finite-size analysis}
\begin{figure}[t]
\centering		
\includegraphics[width=3.265in] {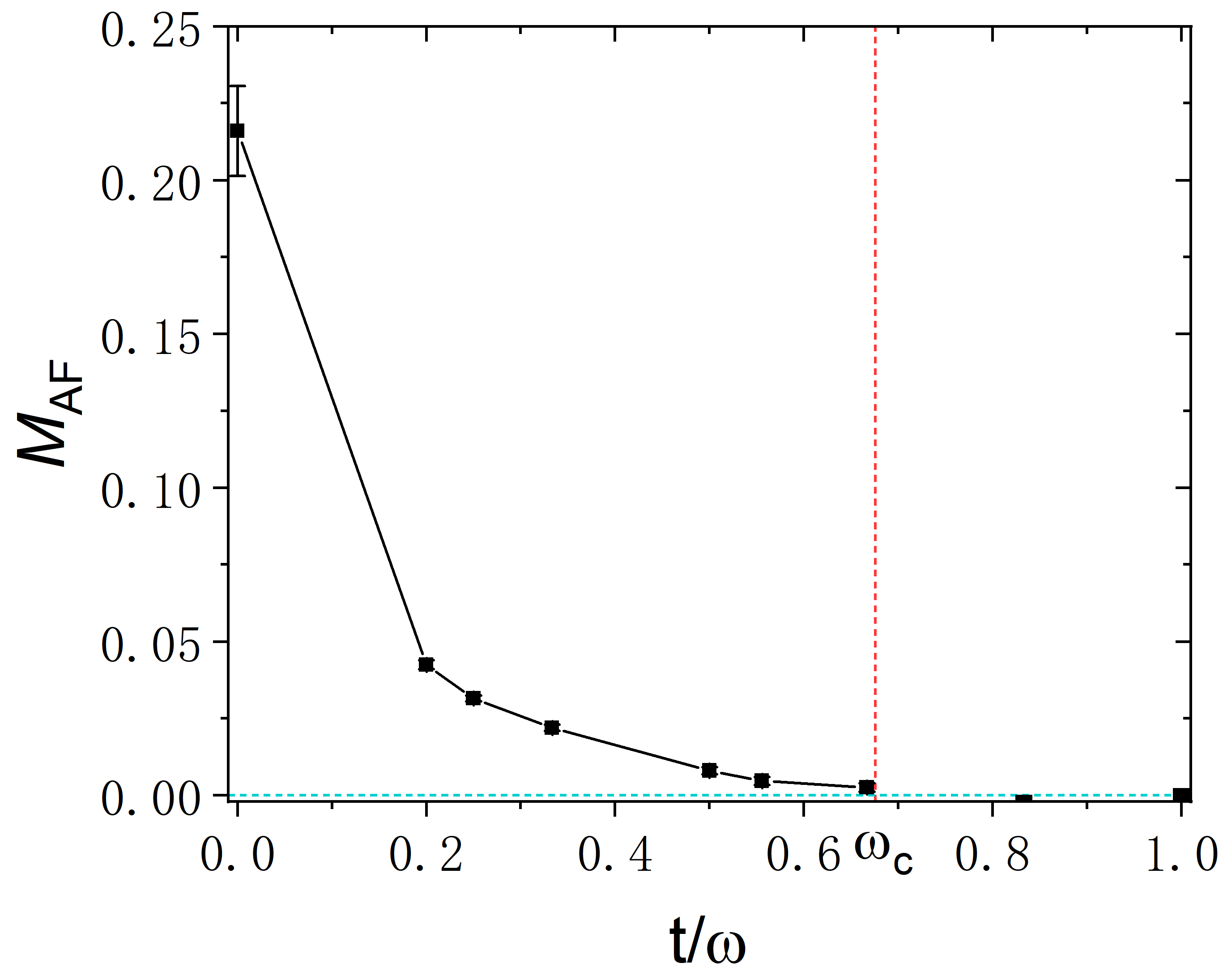}
\caption{The QMC-computed AF order parameter as a function of inverse phonon frequency at $\lambda\approx 0.25$ ($g=1.4$). The red dashed line represents value of corresponding critical frequency discussed in the main text.}
\label{FigOrderParameterg14}
\end{figure}

In the main text, we performed the finite-size scaling to obtain the quantities in the thermodynamic limit. The scaling function we usually use is $C(L)=C(L\ra\infty)+a/L^b$, where $C(L)$ stands for the quantity such as AF order parameter and spin gap computed for system size $L$. We estimate the order parameter in finite lattice as the square root of structure factor $M_\text{AF}=\sqrt{3\,S_\text{AF}(\pi,\pi)}$, where factor 3 comes from the equal contribution of $x,y,z$ components of the spin structure factor. In \Fig{figJAFMorder}, we presented the result of AF order parameters as a function of $\lambda$ in the anti-adiabatic limit. For $\lambda\approx 0.25$ and finite phonon frequency ($0<\omega<\infty$), the extrapolated AF order as a function of $\omega$ is shown in \Fig{FigOrderParameterg14}. The value at $t/\omega=0$ comes from the computation at anti-adiabatic limit. The critical frequency $\omega_c$ discussed in the main text is the phonon frequency where $M_\text{AF}$ vanishes.
	
We compute spin gap $\Delta_s$ according to the asymptotic scaling behavior of time-dependent spin-spin correlation: $\avg{S^z(\v{Q},\tau)S^z(\v{Q},0)}\sim\E{-\Delta_s\tau}$, where $\v{Q}=\inc{\pi,\pi}$. The spin gap $\Delta_s(L)$ for different lattice size $L$ can be extracted by fitting the correlation when $\tau$ is sufficiently large. Then, $\Delta(L)$ is extrapolated to $L\to \infty$ via a power-law scaling function $\Delta(L)=\Delta+a/L^b$ to obtain the spin gap in the thermodynamic limit.

\subsection{C. Exact solution in the adiabatic limit}
In the adiabatic limit ($\omega=0$), phonon displacements $X_{ij}$ become classical quantities without quantum dynamics.  At zero-temperature, the phonon displacements are static and the fermions are described by a quadratic Hamiltonian depending on the phonon displacement $X=\{X_{ij}\}$. Consequently, an exact solution of the ground state of the SSH model in the adiabatic limit can be obtained in the variational sense. In the adiabatic limit, the SSH model \Eq{EqOriginalModel} is reduced to the following quadratic form depending on the phonon configuration $X$:
\beq\label{EqMeanfieldHamiltonian}
H\Inc{X}=\sum_{i,\delta}\left[ \frac{K}{2}X_{i,\delta}^2 +\inc{gX_{i,\delta}-t}(c^\dagger_{i,\s}c_{i+\delta,\s}+h.c.)\right],
\eeq
where $\delta=x,y$ refers to unit vectors on $x$ or $y$ directions and $X_{i,\delta}=X_{i,i+\delta}$ is the phonon displacement on the NN bond $\avg{i,i+\delta}$. The variational method is applied here by minimizing the ground-state energy, which yields the following self-consistent equation: $g\avg{(c^\dagger_{i,\s}c_{i+\delta,\s}+h.c.)}= -K X_{i,\delta}$.

Here we compute the energy assuming various different ansatz of the phonon displacement configuration $X$, including staggered, staircase, columnar, and plaquette VBS patterns. The ordering vector for staggered and staircase patterns is $(\pi,\pi)$ while the ordering vector of columnar and plaquette patterns is $(\pi,0)$ or $(0,\pi)$. Since the Fermi surface of the non-interacting electrons at half filling is perfectly nested by the wave vector $(\pi,\pi)$, it is expected that the VBS patterns with ordering vector $(\pi,\pi)$ is more favored than the ones with $(0,\pi)$ or $(\pi,0)$. Both the staggered and columnar VBS patterns break the lattice $\mathbb{C}_4$ symmetry while the staircase and plaquette ones develop dimer ordering on both $x$ and $y$ directions. We assume $X_{i,\delta}=m_\delta+\inc{-1}^{i_x+i_y} \Delta_\delta^\mathrm{stag}+\inc{-1}^{i_\delta} \Delta_\delta^\mathrm{col}$, 
where $m_\delta$ is the uniform component of the phonon displacements such that the uniform hopping amplitude along $\delta=x,y$ direction is given by $t_\delta=t-gm_\delta$.  
Our calculations clearly show that the SSH phonons in the adiabatic limit favors the staggered VBS state with momentum $\inc{\pi,\pi}$ and $\mathbb{C}_4$ symmetry breaking for any finite EPC constant $\lambda$. Besides, when the staggered VBS order parameter develops spontaneously in $x$-direction, namely $\Delta_x^\mathrm{stag}\neq 0$ while $\Delta_y^\mathrm{stag}=0$, the uniform hopping amplitude $t_x$ and $t_y$ exhibits the anisotropy $t_x>t_y$. Such anisotropy develops so that the shifted Fermi surface $-2t_x\cos{k_x}-2t_y\cos{k_y}=0$ does not cross the nodal lines of gap function $2\Delta_x^\mathrm{stag}\sin{k_x}=0$; namely the Fermi surface can be fully gapped to gain energy.

\subsection{D. The effect of repulsive Hubbard interaction}
\begin{figure}[t]
\includegraphics[width= 3.165in]{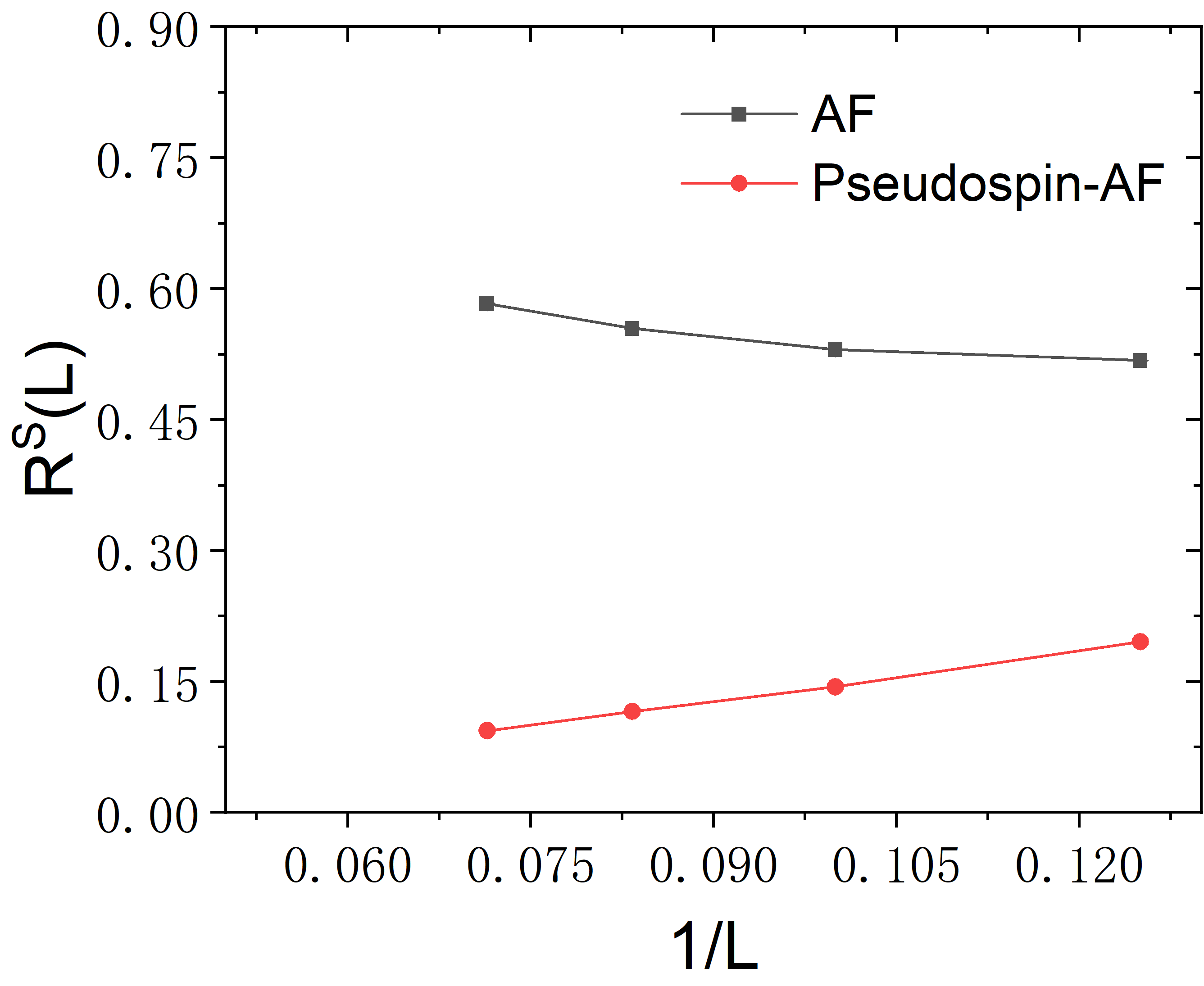}
\caption{The finite size scaling of correlation ratio of AF and pseudospin-AF ordering at $\omega=3t$ and $\lambda\approx 0.25$ ($g=1.4t$) for $U=0.5$. }
\label{FigHubbardU}
\end{figure}

As mentioned in the main text, the SSH model in \Eq{EqOriginalModel} respects the O(4) symmetry, which implies the degeneracy between AF and pseudospin-AF (namely CDW/SC) correlations. However, this degeneracy can be lifted by any finite (even infinitesimal) Hubbard interaction.  
By turning on a weak repulsive Hubbard interaction $H_U$, AF is more favored than pseudospin-AF ordering so that the degeneracy between them is lifted. For instance, in \Fig{FigHubbardU} we present the results of turning on a weak Hubbard $U$ at $\omega=3t$ and $\lambda\approx 0.25$ ($g=1.4t$) where the ground state without $U$ is degenerate between AF and pseudospin-AF (namely CDW/SC) ordering. Once a weak Hubbard $U=0.5$ is turned on, pseudospin-AF (CDW/SC) order is suppressed to zero and AF is the only order in the ground state, as shown in \Fig{FigHubbardU}.   
Thus, in the presence of any weak repulsive Hubbard interaction, AF is the only order when the phonon frequency is larger than a critical value or the EPC constant is less than a critical value.   

\end{document}